\documentclass{article}

\usepackage{arxiv}
\usepackage{authblk}

\usepackage{graphicx}
\usepackage[utf8]{inputenc} 
\usepackage[T1]{fontenc}    
\usepackage{hyperref}       
\usepackage{url}            
\usepackage{booktabs}       
\usepackage{amsfonts}       
\usepackage{nicefrac}       
\usepackage{microtype}      
\usepackage{lipsum}
\usepackage{amsmath}
\usepackage{subcaption}
\usepackage[justification=centering]{caption}
\usepackage[font=small,skip=0pt]{caption}
\usepackage[linesnumbered,boxed ] {algorithm2e}
\usepackage{parskip}
\usepackage{graphicx}
\usepackage[table,xcdraw]{xcolor}
\usepackage{multirow}
\usepackage{caption}
\usepackage{float}
\usepackage{adjustbox}

\usepackage{colortbl}

\SetKwComment{Comment}{$//$\ }{}

\newtheorem{property}{property}
\newtheorem{definition}{Definition}
\newtheorem{proof}{proof}

\newtheorem{theorem}{theorem}
\date{date}
\title{A Neighborhood-preserving Graph Summarization}

\author [1,3]{Abd Errahmane KIOUCHE}
\author [2]{Julien BASTE}
\author [1]{Mohammed HADDAD}
\author [1]{Hamida SEBA}

\affil[1]{Univ Lyon, Université Lyon 1, LIRIS UMR CNRS 5205, F-69621, Lyon, France.\\
 E-mail: \texttt{\{\texttt{abd-errahmane.kiouche,mohammed.haddad,hamida.seba\}@univ-lyon1.fr}}}
\affil[2]{Univ. Lille, CNRS, Centrale Lille,
UMR 9189 - CRIStAL - Centre de Recherche en Informatique Signal
et Automatique de Lille, F-59000 Lille, France. E-mail: \texttt{julien.baste@univ-lille.fr}}
\affil[3]{LCSI, Ecole nationale Supérieure d’Informatique (ESI),Algeria.}

\begin{document}
\maketitle
\date{}
\begin{abstract}

We introduce in this paper a new summarization method for large graphs. Our summarization approach retains only a user-specified proportion of the neighbors of each node in the graph. Our  main aim is to simplify large graphs so that they can be analyzed and processed effectively while preserving as many of the node neighborhood properties as possible. Since many graph algorithms are based on the neighborhood information available for each node, the idea is to produce a smaller graph which can be used to allow these algorithms to handle large graphs and run faster while providing good approximations. Moreover, our compression allows users to control the size of the compressed graph by adjusting the amount of information loss that can be tolerated. The experiments conducted on various real and synthetic graphs show that our compression reduces considerably the size of the graphs. Moreover, we conducted several experiments on the obtained summaries using various graph algorithms and applications, such as node embedding, graph classification and shortest path approximations. The obtained results show  interesting  trade-offs  between the algorithms runtime speed-up and the precision loss.

\end{abstract}

\section{Introduction}

Graphs are widely used in data modeling because of their ability to represent,  in a simple and intuitive way, complex relations and interactions between objects: social interactions, protein-protein interactions, chemical molecule bonds, transport networks, etc. We recall that a graph $G=(V, E)$ is a data modeling tool consisting of  a set $V$ of vertices, also called nodes, and a set $E$ of edges that connect vertices. Vertices represent objects, while edges represent relationships between them. Edges can be directed, and both vertices and edges can have labels. As we are witnessing an explosion in the number of data generated and processed by our applications, it becomes important to deal efficiently with large graphs the processing of which remains a challenging issue. In fact, the amount of generated data is continuously increasing.
Our basic and simple daily activities such as sending emails, surfing websites, purchasing online, and interacting via social networks,  generate, on their own, a huge amount of data each day.
 For example, in 2019 Facebook social network had more than 2.4 Billion monthly active users with an average of 155 friendship links for each user\footnote{https://www.omnicoreagency.com/facebook-statistics/ visited July.2020}.  This large volume of data makes graph querying and analysis a very challenging task. However, a viable solution seems to emerge from the possibilities offered by graph summarizing.\\
Graph summarization, also known by graph compression or simplification, is a solution that tackles scalability and performance issues when dealing with massive graph data. Beyond the reduction of the volume of data which is the main aim of compression, graph summarization looks for significant summaries that can be used, in graph analysis,  without decompression.

In fact, using graph summaries helps to speed-up graph algorithms so that they can efficiently  run on large graphs. Compression algorithms produce smaller graphs or simpler graph representations, which can be maintained in main memory and queried and analyzed in reasonable time.
\\
Many graph algorithms, such as node embedding, node classification, recommendations, shortest path approximation and graph comparison, are based on the neighborhood information available for each node. Finding this information may be difficult in practice as dealing with all the neighbors for each node requires all the edges (links) of the graph to be processed,  which is time and space consuming. This motivated  us to introduce a graph compression that controls the size of the preserved neighborhood of vertices in the computed summary. So, in this paper, we propose a new graph compression which retains only  a user-specified proportion of the neighbors of each node to reduce the size of the graph while preserving neighborhood queries.

The main idea is to sparsify the graph by removing edges, while ensuring that a predefined proportion of the neighbors of each node  is included in the set of $t$-hops neighbors of the node in the compressed graph ($t \geq 1$).

The main advantages of our neighborhood-preserving compression are:
\begin{itemize}
    \item \textbf{Reduction in storage space of the graph: } our compression can decrease drastically the number of edges in the graph, thus allowing the compressed graph to be loaded into main memory. The size of the compressed graph can be controlled by adjusting the proportion of the preserved neighborhood’s information.
    \item \textbf{Fast Approximation of graph algorithms: } Many graph algorithms, such as community detection, shortest path lengths and graph comparison, are mainly based on the neighborhood information of the graph nodes. These algorithms cannot efficiently run on large graphs since they  require all the edges of the original graph to be loaded in main memory.  Since our compression produces a smaller graph that maintains the principal neighborhood’s information, it can be used to allow these algorithms to handle large graphs and run faster while  providing good approximations of the original results.
    \item \textbf{User-controlled trade-off between compression ratio and information loss:} with our compression, the user can control the size of the compressed graph by adjusting the amount of information loss that can be tolerated. This is a very useful property, since the amount of tolerated information loss differs significantly from one application to another, and the desired size of the compressed graph depends mainly on the available memory.

\end{itemize}

The remainder of this paper is organized as follows: Section \ref{Sec-Related} reviews related works on graph compression methods and their applications. Section \ref{Sec-Approach} formally  defines the  problem of neighborhood-preserving graph  compression and studies its complexity. Then, Section \ref{Sec-Algo} provides a description of  the algorithms used to implement this compression. Section \ref{Sec-Results} presents the results obtained through the extensive experiments we undertook to evaluate the compression approach, as well as the usefulness of the obtained summaries. Finally, Section \ref{Sec-Conclusion} concludes the paper and points out some research perspectives.

\section{Related Work}
\label{Sec-Related}
Graph compression is attracting increasing interest in various domains and applications \cite{Liu2018}. The aim of graph compression, considered here, is to compute a graph summary that retains all or part of the original graph properties, thus allowing use of the summary instead of the original graph in certain applications. The obtained summary can be either a graph that is  simpler or smaller than the original graph, or any other data structure that is more compact or is simpler to use than the original graph. Compression algorithms can be classified in three main categories according to how they simplify the input graph: (1) sampling, (2) sparsification, and (3) regularity encoding. Sampling and sparsification based methods generate lossy graph compression and their results are generally a graph.
 Regularity encoding based methods allow having lossy, as well as loss-less, graph summaries, either as graphs or other data structures:
\begin{enumerate}
\item Graph sampling \cite{Bloemena1976,Zhang2017} consists in using a fraction of the graph to make inferences about the whole dataset. It is generally used for dynamic graphs for which a sample at time $t$ is a likely representation of the graph. It is also used with  very large graphs, such as protein to protein interactions, where dealing with the whole graph is too slow. Several graph sampling methods are proposed in the literature. They generally start with a set of initial vertices (and/or edges) which can be empty  and expand the sample based on a specific algorithm such as graph exploration and traversal algorithms. As examples, Breadth-First sampling is used for social network analysis \cite{Ahn2007} and graph mining \cite{Ravkic2018}. In \cite{Yousuf2020}, the authors  apply a traversal based sampling that utilizes only the local information of nodes, combined with estimated values of a set of properties, to guide the sampling process and extract tiny samples that preserve the properties of the graph and closely approximate their distributions in the original graph. Random walks based methods  are also largely applied in large-scale graph analysis \cite{Nazi2015,Li2019,Zhao2019}. Frontier sampling,  an edge
sampling method using multidimensional random walkers,
is used to  estimate the degree distributions and the global clustering coefficient in \cite{Ribeiro2010}.
In \cite{Riondato2016}, the authors approximate betweenness centrality  based on a sampled set of shortest paths.

\item Graph sparsification stands for the methods that compute a sparse subgraph of the input graph, which preserves some of its properties such as cuts or shortest paths \cite{Chew89}. Graph sparsification methods can also rely on sampling as a tool to achieve sparsification.
    Given a social graph and a log of past propagations, the authors of \cite{Mathioudakis2011} prune the network to a prefixed extent, while maximizing the likelihood of generating the
propagation traces in the log. A similar work is described in \cite{Bonchi2013}. It tackles the problem of simplifying a graph, while maintaining the connectivity recorded in a given set of observed activity traces represented by a set of DAGs (or trees) with specified roots. The problem consists in selecting a subset of arcs in the graph so as to maximize the number of nodes reachable in all DAGs by the corresponding DAG roots. This is a cover-maximization problem that the authors bring to a problem of minimizing a submodular function under size constraints and using an algorithm introduced in \cite{Nagano2011} to solve it.
\item Regularity encoding based methods search for regularities within the graph structure, i.e., particular patterns or just repetitive patterns, and then encode these regularities so as to obtain a compact representation of the graph. Several approaches are proposed in the literature and differ by both the kind of considered regularities and how these regularities are encoded within the computed summary.
 Some methods of this class consist in merging or combining similar nodes, or subgraphs into super-nodes, and similar edges into super-edges. Others work directly on the adjacency matrix of the graph using for example $k2$-trees \cite{Brisaboa2009}.
 In \cite{Navlakha2008}, the authors propose a summarizing approach that iteratively aggregates similar nodes, i.e., those that have the greatest number of common neighbors. This aggregation is controlled by an objective function that represents the cost of the compressed output graph and is defined according to the principle of Minimum Description Length (MDL) \cite{Rissanen1978}. The graph is encoded with a summary and a set of correcting edges. These corrections, applied to the summary, enable the initial graph to be reconstructed. Identifying vertices with a similar neighborhood is a well-studied topic known as modular decomposition of graphs \cite{Gallai1967,Habib2010}, which aims to highlight groups of vertices that have the same neighbors outside the group. These subsets of vertices are called \textit{modules}.
  Modular decomposition is used in \cite{Lagraa2016} to compress a graph and compute its exact list of triangles using solely the computed summary.  The compression consists in considering each module as a super-node.
Several works such as  \cite{Suel2001} and \cite{Boldi2004} take advantage of the regularities of the web graph structure, such as locality and similarity properties, to compress its adjacency lists and reduce the number of bits needed to encode a link.
In \cite{Koutra2015,Neil2015}, the authors compress graphs using MDL on a predefined vocabulary of substructures.
In \cite{Maneth2018}, graphs are compressed by recursively detecting repeated substructures and representing them through grammar rules.
  In \cite{Riondato2017}, the authors use a clustering algorithm to partition the original set
of vertices into a number of  clusters, which will be super-nodes
connected by super-edges to form a complete weighted graph.
The super-edge weights are the edge densities between vertices in
the corresponding super-nodes. The goal is to produce a summary
that minimizes the reconstruction error of the original graph.
In \cite{Kumar2018}, the authors merge into super-nodes graph vertices that have common neighbors so that the obtained compression ensures that the efficiency of a given task does not drop below a user-specified threshold. In \cite{Shin2019}, the authors accelerate node grouping using a divide and conquer approach that allows parallel node merging.
In \cite{Fernandes2018}, the authors use tensor decomposition to  group nodes of an evolving graph according to their connectivity patterns into  super-nodes.
In \cite{Amiri2018}, the authors address the problem of preserving node attributes while summarizing diffusion networks. They propose a sub-quadratic parallelizable algorithm that finds the best set of candidate nodes and merges them to construct a smaller network of super-nodes that ensures similar diffusion properties to the original graph.
In \cite{Bloem2020}, the authors use MDL to measure motif relevance based on motif capacity to compress a graph.
In \cite{Kapoor2020}, the authors incrementally compute a summary of an evolving graph using frequent patterns and MDL principle, combined with a set of operations (merge, split, etc.) on the patterns, in order to provide changes that have occurred in the data since the previous state.
\end{enumerate}
In \cite{Lee2020}, the authors hybrid regularity encoding with sparsification by using both node grouping and edge sparsification. This allows optimizing the size of the obtained summary and the graph reconstruction error.

It is interesting to note that few works explore the usefulness of the computed summaries beyond simple neighborhood or reachability queries. Summaries obtained by graph sampling are used to estimate graph parameters and are rarely used as input for graph applications. Most  regularity-encoding based methods do not investigate this issue at all. Graph sparsification methods are generally designed for specific applications because it is difficult to have lossy summaries that can be used in several kind of graph applications. By targeting neighborhood information in our compression and allowing to control the amount of information loss in the computed summary, we aim to be able to use our summaries in a variety of graph applications. In fact, several graph applications, such as node embedding, node classification, recommendations, etc. are based on the availability of node neighborhood information. In the remainder of the paper, we show that controlling the amount of this information in the computed summary allows to reach good trade-off between algorithm speed-up and precision loss when using the summary as input instead of the original graph.

\section{A Neighborhood-preserving graph compression}
In this section, we explore a new graph sparsification method that aims to control the amount of neighborhood information available for each node in the graph. Our goal is to compute a graph summary that can be used instead of the original graph in several applications.
\label{Sec-Approach}
\subsection{Problem Statement}
\label{Sec-Statement}
Let $t \ge 1$ be a positive integer. The main idea of neighborhood-preserving graph compression is to sparsify the input graph by removing edges, while ensuring that, for all $1 \le i \le t$, a proportion $p(i)$ of the neighbors of each node $v$ is included in the set of the $i$-hops
neighbors of $v$ in the  compressed graph. We denote such compression by $(p,t)$-compression where:
\begin{itemize}
\item $p:\mathbb{N}^* \to [0,1]$ is a monotonically increasing function, which represents the proportion of each node's original neighbors that must be retrieved in its $i$-hops neighborhood in the compressed graph.
\item $t$ : is  the minimum value  for which $p$ reaches its maximal value \textit{i.e.},  $p(x)= p(t), \forall x \ge t$.
\end{itemize}

More formally, given an undirected graph $G=(V,E)$, where $V$ is the set of vertices and $E$ is the set of edges, a  $(p,t)$-compression of $G$ is defined as follows:

\begin{definition}
Let $t$ be an integer and $p:\mathbb{N} \to [0,1]$ be a monotonically increasing function satisfying $p(x) = p(t)$, $\forall x>t$. A \emph{$(p,t)$-compression} of a graph $G=(V,E)$ consists in finding a subgraph $G_c=(V_c,E_c)$ of $G$ such that $V_c=V$, $E_c\subseteq E$, and, for each $0 < x \leq t$ and each $v \in V$,  $\left|N_{G}^{1}(v)\cap N_{G_c}^{x}(v)\right| \geq \left|N_{G}^{1}(v)\right|p(x)$, where $N_{G}^{x}(v)$ is the set of all $x$-hop neighbors of $v$ in $G$.
\end{definition}

In other words, the compressed graph $G_{c}$ contains less edges than $G$ and preserves a given amount (equal to $p(t)$) of the original neighbors for each node. In fact, it is required that, for each $0 < x \leq t$, a proportion $p(x)$ of the original neighbors must be accessible within a maximum of $x$ hops in $G_{c}$ using a simple BFS traversal with depth $x$.

\begin{figure}[!h]
\captionsetup[subfigure]{labelformat=empty}
  \begin{center}
  \begin{subfigure}[b]{.40\textwidth}
    \includegraphics[width=1\textwidth, keepaspectratio]{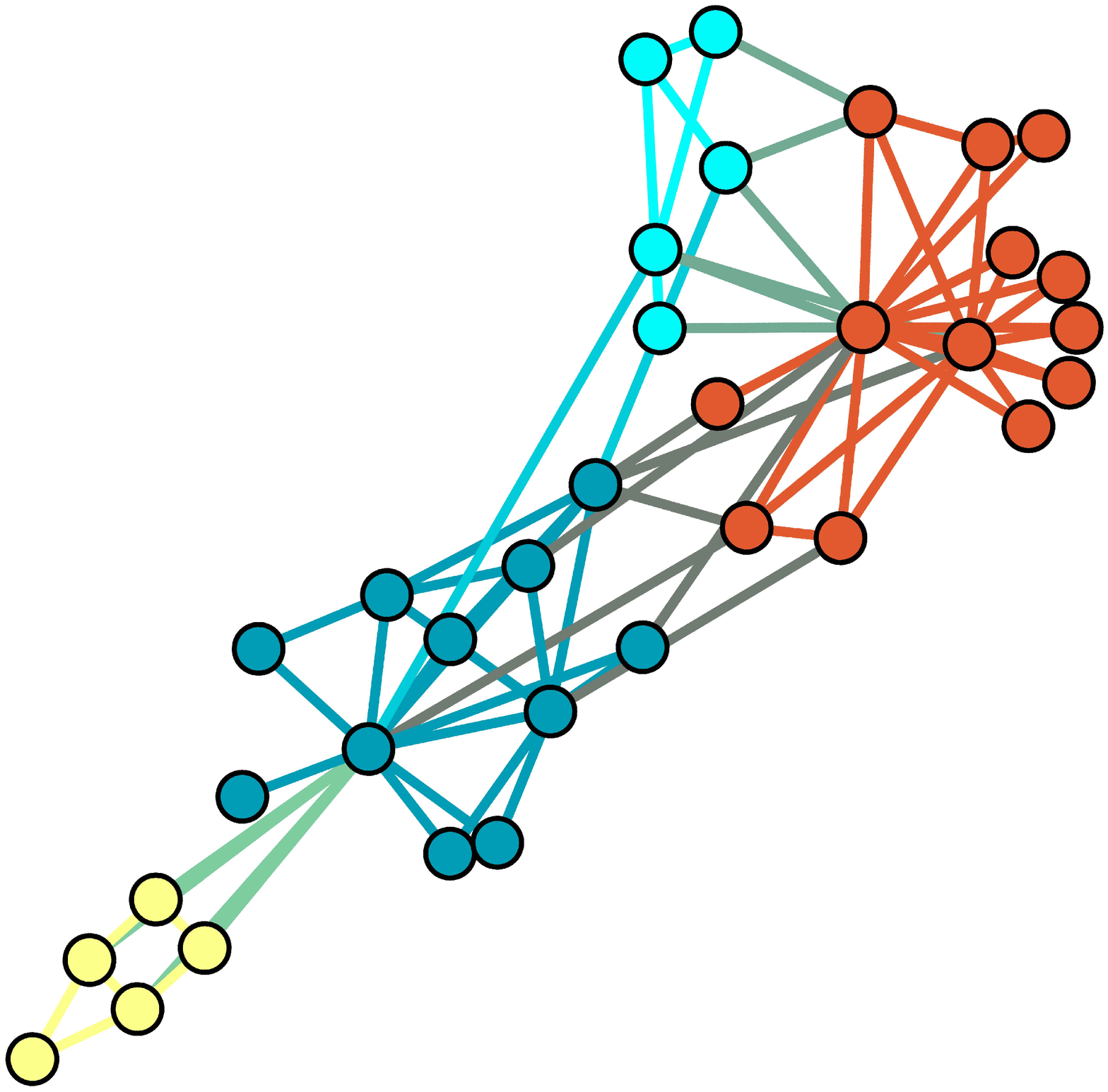}
    \caption{(a) Original graph}
    \label{fig:1}
  \end{subfigure}
  \begin{subfigure}[b]{.50\textwidth}
    \includegraphics[width=1\textwidth, keepaspectratio]{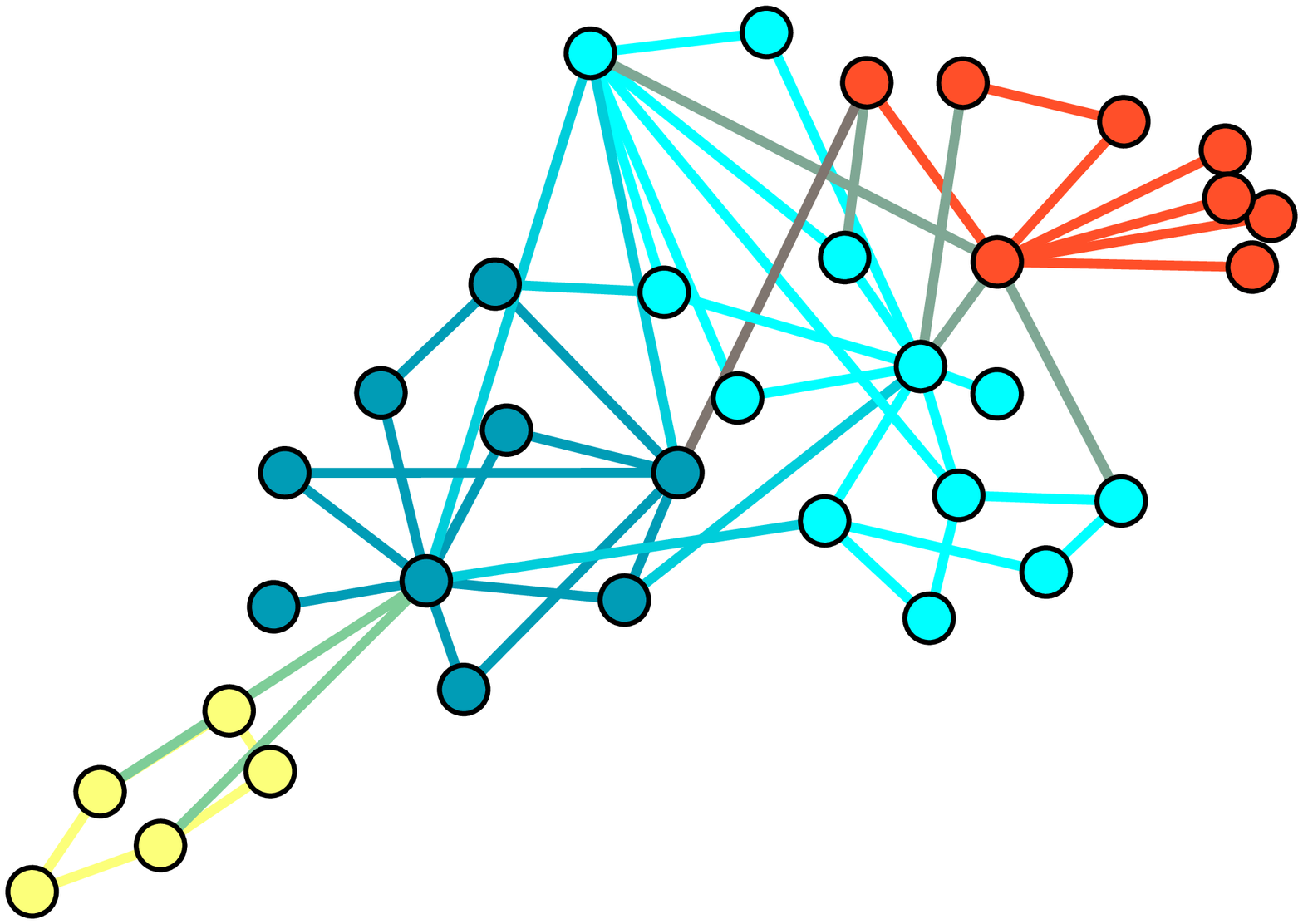}
    \caption{(b) Compressed graph}
    \label{fig:2}
  \end{subfigure}
  \end{center}
  \caption{$(p,2)$-compression of Zachary's karate club network \cite{zachary1977information} where $p(1)=0.5$ and $p(2)=1$ for each node.}
  \label{fig:illustration_zachary}
\end{figure}

Figure \ref{fig:illustration_zachary} illustrates an example of our compression in which $50\% $ of the original neighbors of each vertex are preserved in the compressed graph and are reachable within maximum $2$ hops. The resulting compressed graph is  $30\%$ smaller than the original one.

With $(p,t)$-compression, function $p$ aims to control the loss of neighborhood information at each neighborhood depth. It is obvious that the smaller the preserved proportions, the better the compression ratio, and vice-versa.  As regards parameter t, the higher it is, the bigger the stretch factor of the compressed graph, and vice-versa. The following corollary gives a lower bound of the size of the compressed graph.

\begin{property}
For any $(p,t)$-compression, the number of edges  $|E_{c}|$ of the compressed graph satisfies the following inequality:
$|E|p(1) \leq |E_{c}|$
\end{property}

\begin{proof}
According to the handshaking theorem, we have  $\sum_{v\in V}deg(v)= 2|E|$. Since at least a proportion equal to  $p(1)$ of the original neighbors of each node must be kept in the compressed graph, we have $\sum_{v^{'}\in V_{c}} deg(v^{'})\geq \sum_{v \in V}deg(v)p(1) = 2|E|p(1)$, thus $|E_{c}|\geq p(1)|E|$.
\end{proof}

\begin{figure}[!ht]
\begin{center}
\includegraphics[width=0.7\textwidth, keepaspectratio]{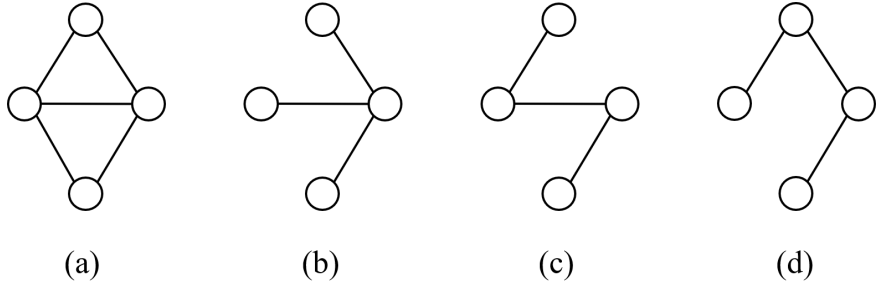}
\end{center}
\caption{a Diamond graph and some of its compressions.}
\label{fig:example_compression}
\end{figure}

It is interesting to note that spanners \cite{Peleg1989} are special cases of $(p,t)$-compression.  Given a graph $G$, possibly edge-weighted, a \emph{graph spanner} (or \emph{spanner} for short) is a subgraph $G'$ which preserves lengths of shortest paths in $G$ up to a multiplicative and/or additive error.  A $t$-spanner is a subgraph $G'$ such that the distance between two vertices in $G'$ is at most  $t$ times the distance between the same two vertices in $G$. Thus, a $t$-spanner is a particular $(p,t)$-compression that could be defined such that $p(i) = 0$ for every positive integer $i < t$ and $p(i) = 1$ for $i \ge t$. In other words, a $t$-spanner is a $(p,t)$-compression whose proportion function $p(x)$ is the Shifted Unit Step Function $u(x-t)$. From a structural point of view, Figure \ref{fig:example_compression} gives an illustration of a 2-spanner of the Diamond graph (Figure \ref{fig:example_compression}.(b)) a 2-spanner of the Diamond graph which is also a $(p,2)$-compression where $p(1) = \frac{1}{2}$ and $p(2) = 1$ (Figure \ref{fig:example_compression}.(c)) and a $(p,2)$-compression where $p(1) = \frac{1}{3}$ and $p(2) = \frac{2}{3}$ which is not a 2-spanner (Figure \ref{fig:example_compression}.(d)).

We can see in this figure that with a $2$-spanner, the left-most vertex is only connected to  $\frac{1}{3}$ of its original neighbors in the Diamond graph (see Figure \ref{fig:example_compression} (b)), while all the vertices keep at least half of their original neighbors with a $(p,t)$-compression (see Figure \ref{fig:example_compression} (c)).

\subsection{NP-hardness and hardness of approximation}
Peleg and Sch\"affer \cite{Peleg1989} showed that, given an unweighted graph $G$, and integers $t \ge 2$, $m \ge 1$, determining if $G$ has a $t$-spanner containing $m$ or fewer edges is NP-complete, even when $t$ is fixed to be $2$. The reduction is from the edge dominating set problem on bipartite graphs. Since a $t$-spanner is a particular $(p,t)$-compression of the graph $G$, we can deduce the following result:

\begin{theorem}
Finding the optimal (smallest) compressed graph satisfying the $(p,t)$-compression constraints for $t\geq2$ is an NP-Hard problem.
\end{theorem}

Another work giving us an alternative proof is Cai's \cite{CAI1994}. He showed that for any fixed $t \ge 2$, the minimum $t$-spanner problem is NP-hard, and for $t \ge 3$, the problem is NP-hard even when restricted to bipartite graphs. The reduction is from the 3-SAT problem.

 Dinitz \textit{et al.} \cite{Dinitz2016} show that for $t \le 3$, and for all $\epsilon > 0$, the $t$-spanner problem cannot be approximated with a ratio better than $2^{(log^{1-\epsilon} n)/k}$ unless $NP \subseteq BPTIME(2^{polylog(n)})$. This implies the same inapproximability result for $(p,t)$-compression.

Concerning the best known approximation ratio, Elkin and Peleg \cite{ELKIN2005} propose approximation algorithms with a sublinear approximation ratio, and study certain classes of graphs for which logarithmic approximation is feasible. It is also shown in \cite{KORTSARZ1994} that for $t = 2$, the $t$-spanner problem admits an $O(log(n))$ approximation. All these results strongly indicate that finding better or even equivalent approximations for the $(p,t)$-compression problem will be a hard task. In particular, finding a better result should begin with finding better approximation than $O(log(n))$ for the $t$-spanner problem.
\subsection{Integer linear programming formulation}
\label{Sec-LP}
In the following, we provide an integer linear programming formulation of our problem.

Given an input graph $G = (V,E)$,
we denote by $\mathcal{W}$ the set of all paths in $G$.
Given $e = \{u,v\} \in E$, we denote by $\mathcal{W}_{uv}$ the set of all paths in $\mathcal{W}$ from $u$ to $v$.
Note that the graph is undirected and so $\mathcal{W}_{uv}$  also corresponds to the paths from $v$ to $u$.
Moreover, given $i \in \mathbb{N}$ we denote by $\mathcal{W}_{uv}^i$ the set of all paths of $\mathcal{W}_{uv}$ of size at most $i$.

We then define the used variables:
\begin{itemize}
\item $x_e$, for each $e \in E$, denotes whether $e \in E$ is selected ($x_e = 1$) or deleted ($x_e = 0$).
\item $f_w$, for each $w \in \mathcal{W}$, is such that $f_w = 0$ if at least one edge $e$ of $w$ is such that $x_e = 0$.
  Note that we can have a path $w \in \mathcal{W}$ such that every edge $e$ of the path is such that $x_e = 1$ but still have $f_w = 0$.
\end{itemize}

We can now write the integer linear programming equation:
\begin{align}
  \label{eq:lp1}
  \min\quad &  \sum_{e\in E}x_e&  &\\
  \label{eq:lp2}
  \text{s.t.\quad}  &  f_w \leq x_e&  &\forall w\in \mathcal{W}, e\in E:~ e \in w \\
  \label{eq:lp3}
            &  \sum_{w\in \mathcal{W}_{uv}} f_w \leq 1&  &\forall uv\in E, i\in \mathbb{N}\\
  \label{eq:lp4}
             &  \sum_{v\in N(u)} \sum_{w\in \mathcal{W}_{uv}^i}f_w \geq p(i) |N(u)| &  &\forall u \in V, i \in \mathbb{N}\\
  \label{eq:lp5}
             &  x_e \in \{0,1\} &  &e \in E\\
  \label{eq:lp6}
             &  f_w \in \{0,1\} &  &w \in \mathcal{W}
\end{align}

The variable $f_w$ can be seen as a flow from the source to the sink.
(\ref{eq:lp2}) ensures that if a path $w \in \mathcal{W}$ uses a removed edge $e$ (i.e., $e$ is such that $x_e = 0$), then the flow $f_w = 0$.
Using (\ref{eq:lp3}), we know that for each $uv \in E$ there exists at most one path $w \in \mathcal{W}_{uv}$ such that $f_w = 1$.
By intuition, we assume that we took the path $w \in \mathcal{W}_{uv}$ of shortest length that is still available in the remaining graph after removing the edges $e\in E$ such that $x_e = 0$.
Then condition (\ref{eq:lp4}) ensures that the number of neighbors of a vertex $u$, which are now at distance at most $i$ in the new graph, is at least the proportion given by $p(i)$.


\section{Algorithms and approximations}
\label{Sec-Algo}
In this section, we present four algorithms for finding the $(p,t)$-compression of an input graph $G$.
Since finding the optimal $(p,t)$-compression is NP-Hard and cannot be resolved in polynomial time, we propose polynomial time approximations (sub-optimal algorithms). Algorithm \ref{alg:Algo1} gives the basic implementation of our compression. It has the advantage of simplicity and speed.
Algorithm \ref{alg:Algo1} takes as input a simple graph to compress $G=(V,E)$, the compression parameters $p$ and $t$ and an order $E_o$ for processing the edges of the input graph. $E_o$ is by default a random ordering of the vertices. The idea is to process the edges of the initial graph in the order $E_o$. The algorithm processes  the edges of $G$ incrementally as follows:
If an edge $e$ can be removed from $G$  without violating the neighborhood preservation constraints, the algorithm does not keep this edge in the summary. Otherwise, the algorithm keeps the edge in the summary. Assume that the average branching factor ( Out/In degree) of the graph is equal to $b$, then average time complexity of Algorithm \ref{alg:Algo1} is $O(|E|b^t)$.

\IncMargin{1.5em}
\begin{algorithm}[t]
\SetAlgoLined
\KwData {
         $G = (V,E)$ a simple Graph, \\
        \hspace{11mm} $t$ an integer, \\
        \hspace{11mm} $p$ a monotonically increasing function $p:\mathbb{N} \to [0,1]$, \\
        \hspace{11mm} $E_o$ a possible order of the graph edges}
\KwResult{$G_c = (V_c,E_c)$ a compressed graph }
 \Comment*[h]{Initialization Step }\;
 $G_c=(V_c,E_c) \gets (V, \emptyset)$\;
 $G'=(V',E') \gets (V, \emptyset)$\;
 \For{$e=(u,v) \in E_o$} {
 $ E' \gets E' \cup \{ (u,v)\}$\;
 $N_{G'}^{1}(u) \gets $ direct neighbors of node $u$ in $G'$\;
 $N_{G'}^{1}(v)\gets $ direct neighbors of node $v$ in $G'$\;
 insert $ \gets False$\;
 \For{$i=1$ to $ t$} {
    $N_{Gc}^{i}(u) \gets $  neighbors of node $u$ in $G_c$ within at most $i$-hops\;
    $N_{Gc}^{i}(v) \gets $  neighbors of node $v$ in graph $G_c$ within at most $i$-hops\;
    \If{$ |N_{Gc}^{i}(u) \cap N_{G'}{1}(u)| < p(i)|N_{G'}^{1}(u)|$ or $|N_{Gc}^{i}(v) \cap N_{G'}^{1}(v)| < p(i)|N_{G'}^{1}(v)|$}{
       insert $\gets True$\;
       \textbf{Break}\;
    }
 }
 \If{ insert  } {
     $ E_c \gets E_c \cup \{ (u,v)\}$\;
 }
 }
 \caption{Basic Algorithm}
 \label{alg:Algo1}
\end{algorithm}
\DecMargin{1.5em}

We note that different edge orderings lead to different compression performances. Therefore, in order to  improve the compression performance of our algorithm, we propose in the three following subsections, 3 sub-optimal algorithms which are based on the basic algorithm and try to find a near optimal edge processing order.

\subsection{Linear programming}

We provide in Section \ref{Sec-LP}, an optimal integer linear programming formulation of $(p,t)$-compression. However, such a resolution is \textsf{NP}-hard to solve, so we use the standard tricks consisting in relaxing the problem. For this, we keep the same formulation but allow the values  $x_e$, $e\in E$, and $f_w$, $w \in \mathcal{W}$, to be any real values between $0$ and $1$. As we only have to consider paths of a length at most $t$, we have a polynomial number of variables  (the degree of which depends on the fixed value $t$). The average number of variables is of the order $O(|E|+|V|b^t)$, where $b$ is the average branching factor of the graph. This resolution provides a value for each $x_e$, $e \in E$. The interpretation we give to this resolution is that the higher the value of $x_e$, the more likely we want to keep $e$ in our solution. In reverse, the lower the value of $x_e$, the more likely we want to remove the edge $e$. Thus, we can use the values of $x_e$, $e \in E$, in order to obtain an ordering for the edges and give this ordering to the basic compression algorithm (see Algorithm \ref{alg:Algo2}). Since this linear problem is solvable in polynomial time, the time complexity of  Algorithm \ref{alg:Algo2} is $O(poly(|E|+|V|b^t))$.

\IncMargin{1.5em}
\begin{algorithm}[th]
\SetAlgoLined
\KwData {
         $G = (V,E)$ a simple Graph, \\
        \hspace{11mm} $t$ an integer, \\
        \hspace{11mm} $p$ a monotonically increasing function $p:\mathbb{N} \to [0,1]$}
\KwResult{$G_c = (V_c,E_c)$ a compressed graph }
 \Comment*[h]{Computing the greedy edge order $E_{go}$  }\;

  Solve the LP Relaxed problem to compute the edge scores $x_e$\;

  $E_{go} \gets $ sort edges E in descending order according to their score $x_e$\;

  $G_{c} \gets$ Basic Algorithm ($G$,$t$,$p$,$E_{go}$)\;
 \caption{LP Algorithm}
 \label{alg:Algo2}
\end{algorithm}
\DecMargin{1.5em}

\subsection{Greedy order based on edge connectivity }
Computing the LP-based order is time-consuming for large graphs according to its time complexity. So, we propose in this subsection another edge ordering that can be computed much faster than the LP order. The idea is to first process the edges with a high centrality value.
The centrality we consider here is a relaxation of local edge betweenness defined in \cite{Gregory2008}. An edge with a high edge betweenness centrality represents a bridge-like connector between two parts of a network, the removal of which may affect the shortest paths between them. The local edge betweenness of an edge $e$ is the number of shortest paths running along $e$, the length of which is less than or equal to some constant $t$. In our relaxation, we consider all simple paths of a length at most $t$, i.e., not necessarily shortest paths. Thus, we compute for every edge $e$ a centrality score $s(e)$ according to Equation \ref{equation_7}.

In Equation \ref{equation_7}, $\sigma_t(u,v |e)$ is the number of simple paths from $u$ to $v$ of length $\leq t$ that pass through the edge $e$. Once all scores are computed, we sort the edges in descending order according to their score $s(e)$ and pass the obtained order as input to the basic algorithm (See Algorithm  \ref{alg:Algo3}). The average time complexity of Algorithm  \ref{alg:Algo3} is $O((|E|+|V|b^t )log (|E|+|V|b^t))$.
\begin{equation}
\label{equation_7}
  s(e) = \sum_{(u,v)\in E} \sigma_t(u,v | e) \ \forall (u,v) \in E
\end{equation}

\IncMargin{1.5em}
\begin{algorithm}[th]
\SetAlgoLined
\KwData {
         $G = (V,E)$ a simple Graph, \\
        \hspace{11mm} $t$ an integer, \\
        \hspace{11mm} $p$ a monotonically increasing function $p:\mathbb{N} \to [0,1]$}

\KwResult{$G_c = (V_c,E_c)$ a compressed graph }
 \Comment*[h]{Computing the greedy edge order $E_{go}$  }\;

  \For{$e \in E$} {
    compute the score s(e) using Equation \ref{equation_7}\;
  }

  $E_{go} \gets $ sort the edges of $G$ in descending order according to their score $s(e)$\;

  $G_{c} \gets$ Basic Algorithm ($G$,$t$,$p$,$E_{go}$)\;
 \caption{ Greedy Algorithm based on edge connectivity (EC)}
  \label{alg:Algo3}
\end{algorithm}
\DecMargin{1.5em}

\subsection{Sub-optimal order based on Simulated Annealing}

In the previous two subsections, we have proposed two greedy edge orderings to improve compression performance. However, the drawback of these two solutions is that they are more time-consumingthan Algorithm \ref{alg:Algo1} with a random edge ordering, as we will reveal in the next section with the experimental evaluation. Moreover, the computation time cannot be controlled by the user since the computation of both orderings, i.e., the LP ordering and the greedy ordering based on edge connectivity, cannot  be suspended,  and we need to go to the end of the calculation. To overcome this problem, we propose a third algorithm based on Simulated Annealing (SA) \cite{van1987simulated}. The advantage of this solution is that the computation time can be controlled by the user by adjusting the number of SA iterations. SA is an optimization scheme that allows efficient search space exploration by accepting, with a given probability, worst solutions to avoid a premature convergence \cite{van1987simulated}. SA for $(p,t)$-compression is illustrated in Algorithm \ref{alg:Algo4}. The initial state of the  algorithm  is a random order of edges. Then, in each iteration, the algorithm makes a slight modification to edge order by performing two permutations of two elements in the vector representing the order and recomputes the cost of the new solution. If the new order is better, the algorithm keeps the order. Otherwise, the algorithm keeps it with  a probability which increases over the iterations (see line 19 of Algorithm \ref{alg:Algo4}). 

\IncMargin{1.5em}
\begin{algorithm}[th]
\caption{$(p,t)$ Compression  based on simulated annealing}
 \label{alg:Algo4}
\SetAlgoLined
\KwData {
         $G = (V,E)$ a simple Graph, \\
        \hspace{11mm} $t$ an integer, \\
        \hspace{11mm} $p$ a monotonically increasing function $p:\mathbb{N} \to [0,1]$,\\
        \hspace{11mm} $N$ an integer (Number of iterations), \\
        \hspace{11mm} $T_0$ a double ( Initial temperature), \\
        \hspace{11mm} $\alpha$ a double ( decreasing factor)}
\KwResult{$G_c = (V_c,E_c)$ a compressed graph }

  $S \gets$ Random order of $E$\;
  $T \gets T_0$\;
  $G_t(V_t, E_t) \gets $  Basic Algorithm( $G$,$t$,$p$,$S$)\;
  $C_{best} \gets |E_t|$\;
  $C_{S} \gets |E_t|$\;
  \For{$i=1$ to $N$} {
    $S_2 \gets$ Perturbing $S$ by swapping the order of two random edges\;
    $G_t(V_t, E_t) \gets $ Basic Algorithm($G$,$t$,$p$,$S_2$)\;

    \If{$ |E_t| < C_{best} $}{
       $ E_{best} \gets  S$\;
       $ C_{best} \gets |E_t|$\;
    }
    \If{$ |E_t| < COST_{S} $}{
       $ S \gets S_2$\;
       $ C_{S} \gets |E_t|$\;
    }
    \Else{
        $r \gets $ random number between $0$ and $1$\;
        \If{$ \exp(\frac{C_S - |E_t|}{T}) > r $}{
              $ S \gets S_{2}$\;
             $ COST_{S} \gets |E_t|$\;
        }
    }
    $ T \gets \alpha*T$;
  }
  $G_{c} \gets $  Basic Algorithm ($G$,$t$,$p$,$E_{best}$)\;
 \end{algorithm}
\DecMargin{1.5em}

\section{Experimental Analysis}
\label{Sec-Results}
In this section, we present an experimental analysis of our compression. First, we  evaluate the approximation algorithms provided to compute the compression. Then, we provide an analysis of the sensitivity of the compression to parameters $p$ and $t$. Finally, we evaluate its effectiveness on several tasks  such as graph properties estimation, node embedding  and whole graph embedding. All the experiments are carried out on  an Intel core $i7$ processor with $64$ Gigabytes of memory.

\subsection{Evaluation of approximation algorithms }

In this subsection, we present a comparative experimental study of the four proposed approximations of $(p, t)$-compression. For this, we launched the 4 approximations, i.e., the basic algorithm with random edge ordering (Algorithm~\ref{alg:Algo1}), the LP approximation (Algorithm~\ref{alg:Algo2}), the EC approximation (Algorithm~\ref{alg:Algo3}), and the SA approximation (Algorithm~\ref{alg:Algo4}), on 3 families of synthetic graphs the properties of which are given in Table~\ref{tab:dat_syn}.

\begin{table}[ht]
\centering
\caption{Characteristics of the synthetic graph families}
\begin{tabular}[t]{lccc}
\hline
Name& number of graphs  & $|V|$ & $|E|$\\
\hline
SYNTHETIC 1     & $30$        & $20$     & $60$      \\
SYNTHETIC 2     & $30$        & $50$     & $350$     \\
SYNTHETIC 3     & $30$        & $100$     & $1.4K$   \\
\hline
\end{tabular}
\label{tab:dat_syn}
\end{table}

We use the following compression parameters $t=2$ , $p(1)=0.0$ and $p(2)=0.5$. For a reliable and accurate comparison, we carried out around thirty tests on each family of graphs for each algorithm. The results of the comparison are depicted in Table~\ref{tab:comp_approx}.
Note that the  user configuration of the SA is $T_0=10$, $N=1000$ and $\alpha=0.99$. We notice that the two greedy algorithms LP and EC, and the SA algorithm outperform the basic algorithm with a random ordering of edges in terms of compression performance. The results clearly show that the greedy (EC) and the SA algorithms are the best algorithms. The greedy (EC) algorithm seems really interesting and offers the best trade-off between compression performance and runtime. However, all the approximations are still much slower than the basic algorithm with a random order of edges. Therefore, we recommend using the basic algorithm for large graphs.

\begin{table}[ht]
\centering
\caption{Evaluation of the approximation algorithms}
\begin{tabular}{|c|cc|cc|cc|cc|}
\hline
\multirow{2}{*}{Dataset} & \multicolumn{2}{c|}{Basic}  & \multicolumn{2}{c|}{Greedy ( LP)} & \multicolumn{2}{c|}{Greedy ( EC)} & \multicolumn{2}{c|}{SA} \\ \cline{2-9}
                         & Avg $|E_c|$ & time           & Avg $|E_c|$        & time         & Avg $|E_c|$            & time      & Avg $|E_c|$       & time \\ \hline
SYNTHETIC 1              & 28        & \textbf{0.001} & 25.24             & 0.02         & 23.55                & 0.01      & \textbf{21.56}  & 0.5  \\
SYNTHETIC 2              & 121.63    & \textbf{0.008} & 113.26            & 2.5          & 105.66               & 0.02      & \textbf{105.9}  & 5.2  \\
SYNTHETIC 3              & 367.03    & \textbf{0.05}  & 354.36            & 212          & \textbf{323.23}      & 0.09      & 340.4           & 40   \\ \hline
\end{tabular}
\label{tab:comp_approx}
\end{table}

\subsection{Impact of the compression parameters}
In this series of experiments, we study the  effect of parameters $p$ and $t$ on compression performance. To this end, we evaluate our compression using two metrics: the compression runtime measured in seconds and the compression ratio that represents the ratio of the number of deleted edges over the total number of edges (see Equation \ref{Eq_spacegain}).
\begin{equation}
\label{Eq_spacegain}
compression\; ratio  =  \frac{|E|-|E^{'}|}{|E|}
\end{equation}

Note that higher is the compression ratio better is the storage space gain ensured by the compression.
For these experiments and all the following ones we use real graph datasets. Table \ref{tab:datasets} gives the main characteristics of these datasets.

\begin{table}[ht]
\centering
\caption{Characteristics of the real datasets used in our experiments}
\begin{tabular}[t]{lccc}
\hline
Name& number of graphs  & $|V|$ & $|E|$\\
\hline
BLOG-CATALOG    & $1$        & $10.31K$  & $333.98K$ \\
CA-ASTROPH      & $1$        & $18.77K$  & $198.11K$ \\
CA-HEPTH        & $1$        &	$9.8K$	 & $25.9K$\\
COLLAB        & $5000$     & $372.5K$  & $49.1M$   \\
ENZYMES       & $600$      & $19.5K$   & $74.6K$   \\
FLICKR        & $1$        & $80.51K$  & $5.89M$   \\
PROTEINS      & $1113$     & $43.5K$   & $162.1K$  \\
\hline
\end{tabular}
\label{tab:datasets}
\end{table}

Table \ref{tab:tradeoff} gives the compression ratio obtained by our compression on the CA-AstroPh dataset, while varying the neighborhood preservation proportion $p$. As expected, the compression ratio decreases as the preserved proportion of neighborhood increases and vice-versa. Most of the values of the compression ratio obtained with  the various combinations of parameters are satisfactory. In addition, we remark that the compression ratio range is wide (from $7\%$ to $75\%$) which confirms the possibility of controlling effectively the trade-off  information loss/compression ratio using parameters $p$ and $t$. Furthermore, we set $p(t) =1$ in all experiments, which means that the whole  initial neighborhood of each node can be retrieved in a neighborhood of radius $r=t$ at maximum. This ensures that reachability queries are fully preserved for all vertices.  The choice of the best combination of parameters depends essentially on the nature of the graph to be compressed and the user needs. Particularly, for this example, the combinations $(0.5,1)$ and $(0.7,1)$ seem really interesting.

\begin{table}[ht]
\centering

\caption{Compression ratio of the Ca-AstroPh dataset with different combinations of parameters $p$ and $t$}
\begin{tabular}{ccccc}
\hline
\multicolumn{1}{l}{$t$} & $p(1)$ & $p(2)$ & $p(3)$ & compression ratio \\ \hline
\multirow{4}{*}{$2$}    & $0.2$  & $1.0$    & -    & $58.13\%$          \\
                      & $0.5$  & $1.0$    & -    & $45.82\%$          \\
                      & $0.7$  & $1.0$    & -    & $26.39\%$          \\
                      & $0.9$  & $1.0$    & -    & $7.43\%$           \\ \hline
\multirow{4}{*}{$3$}    & $0.0$    & $0.2$  & $1.0$    & $75.00\%$          \\
                      & $0.2$  & $0.5$  & $1.0$& $71.50\%$          \\
                      & $0.5$  & $0.7$  & $1.0$    & $46.73\%$         \\
                      & $0.7$  & $0.9$  & $1.0$    & $26.43\%$          \\ \hline
\end{tabular}
\label{tab:tradeoff}
\end{table}

\begin{figure}[ht]
    \centering
     \includegraphics[width=0.6\textwidth, keepaspectratio]{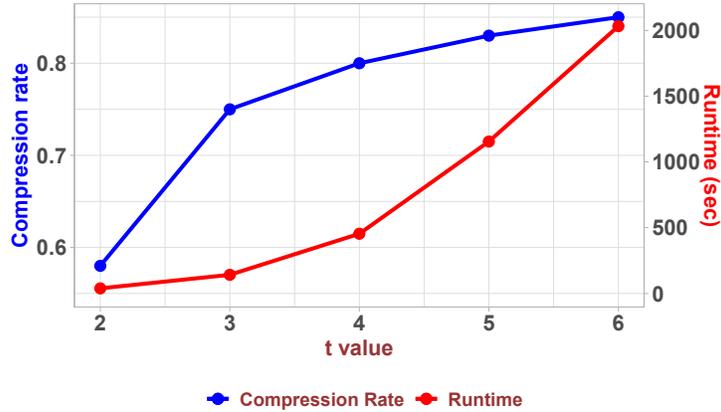}
     \caption{Compression performance of t-spanners on the Ca-AstroPh dataset}
     \label{fig:compressionVStime}
\end{figure}

The curves depicted in Figure \ref{fig:compressionVStime} show the runtime and the compression ratio as a function of the value of  $t$ where $p(0<x<t)=0$ and $p(t)=1$. Note that this combination of parameters gives a particular type of subgraphs called $t$-spanners \cite{Peleg1989}. We notice that the compression ratio grows slowly and starts to level off from  $t=5$. However, the execution time increases exponentially and rapidly. This is due to the complexity of the compression, which is of  the order $O(|E|b^t)$ in the average case. Although spanners give good compression ratios on this dataset ranging from $58\%$ up to $85\%$, they do not allow good control of the trade-off between information loss and compression ratio. Indeed, for this dataset, spanners give a control margin of $(85\%-58\%=27\%)$ for a maximal stretch factor $t=6$, which represents a significant loss of neighborhood information. However, with $(p,t)$-compression we get a larger control margin of $(75\%-8\%= 67\%)$ with a maximal stretch factor $t=3$. This confirms once again the efficiency and usefulness of our compression and its parameter  $p$ when compared to spanners.

\subsection{Applications of \texorpdfstring{$(p,t)$}-compression}
Several graph algorithms are  based on the availability of neighborhood information of nodes. Our first motivation is to be able to use these kinds of algorithms directly on the compressed graphs. So, the purpose of these experiments is to show the effectiveness of our compression in terms of speeding-up for such graph algorithms, while handling large graphs and providing good approximations of the original results. For this, and for all the following experiments, we use the datasets presented in Table \ref{tab:datasets} and compute two new metrics in addition to the compression ratio namely:
\begin{itemize}
    \item \textbf{Speed-up factor:} the ratio between the algorithm run-time on the original graph and its run-time on the compressed graph. he higher the speed-up factor, the faster the graph algorithm on the compressed graph.
    \item \textbf{Performance loss:} the difference between the performance metric value on the original graph and the performance metric value on the compressed graph. For example, for a classification task the performance loss is the difference between the accuracy on the original graph and the accuracy on the compressed graph.The smaller the performance loss, the better the approximation of the graph properties on the compressed graph.
\end{itemize}

\subsubsection{Shortest paths approximation}
The most suitable application for our compression is the approximation of the shortest paths between all nodes, since every $(p,t)$-compression where $p(t)= 1.0$ preserves all the connectivity properties between the nodes of the graph, by stretching all the connecting paths  by a factor equal to $t$ in the worst case. In this experimental phase, we compressed three unweighted undirected graphs with  the following combination of parameters $t=2$, $p(1)=0.5$, and $p(2)=1.0$. Then, we applied the BFS (Breadth First Search) algorithm to compute all shortest paths between all nodes. Table \ref{tab:SPsup} summarizes the obtained compression ratio and the shortest path speed-up obtained on three datasets: CA-ASTROPH, CA-HEPTH and BLOG-CATALOG. We notice that our compression saves a considerable amount of storage space  (compression ratio $> 31\%$) while approximating faster (speed-up ranges from $1.06$ to $1.49$) the shortest path lengths for the 3 chosen datasets. Indeed, reducing the number of edges of the graphs reduces the run-time of the shortest paths computed by the BFS algorithm, which is of complexity $O(|V|(|E|+|V|))$. This speed-up is more noticeable for denser graphs.

\begin{table}[ht]
\centering
\caption{Speeding-up all shortest paths computation}
\begin{tabular}{ccc}
\hline
Dataset      & Space gain & Speed-up \\ \hline
CA-ASTROPH    & $45.82\%$                       & $1.496$    \\
BLOG-CATALOG  & $46.52\%$                       & $1.323$    \\
CA-HEPTH    & $31.08\%$                       & $1.065$    \\ \hline
\end{tabular}
\label{tab:SPsup}
\end{table}

\begin{figure}[ht]
\captionsetup[subfigure]{labelformat=empty}
  \begin{center}
  \begin{subfigure}[b]{.49\textwidth}
    \includegraphics[width=1\textwidth, keepaspectratio]{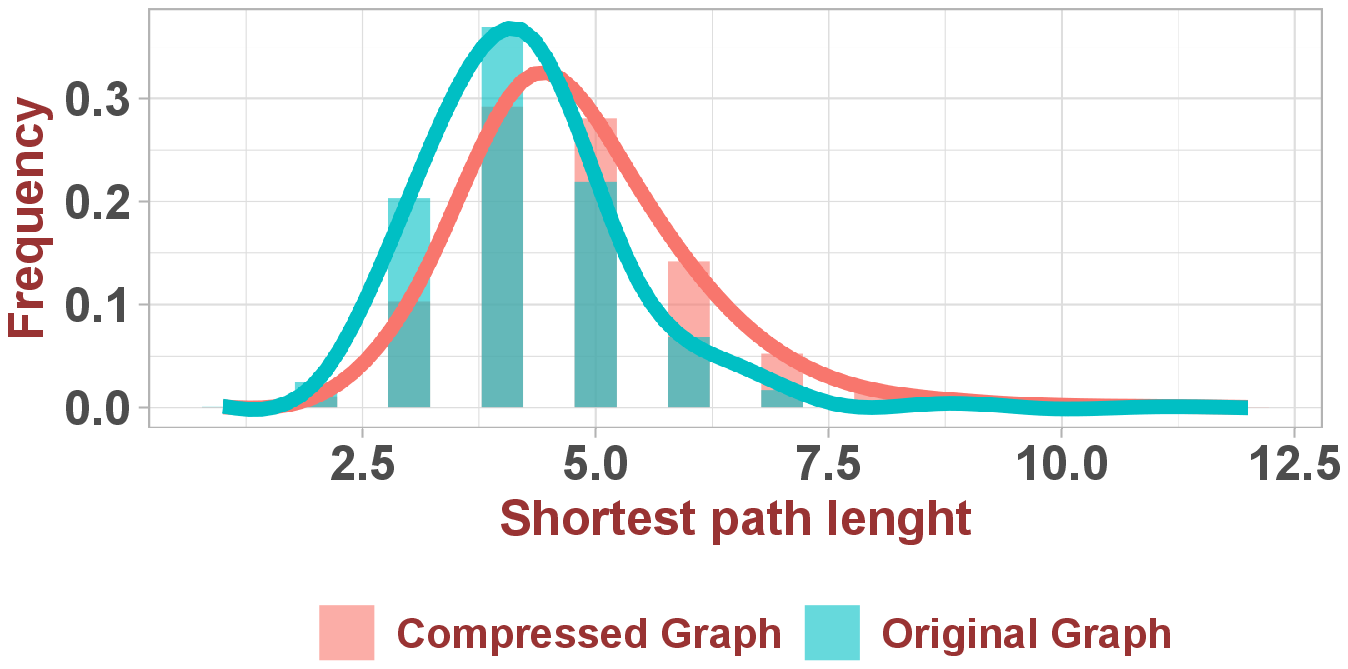}
    \caption{(a) Ca-AstroPh dataset}
    \label{fig:3-a}
  \end{subfigure}
  \begin{subfigure}[b]{.49\textwidth}
    \includegraphics[width=1\textwidth, keepaspectratio]{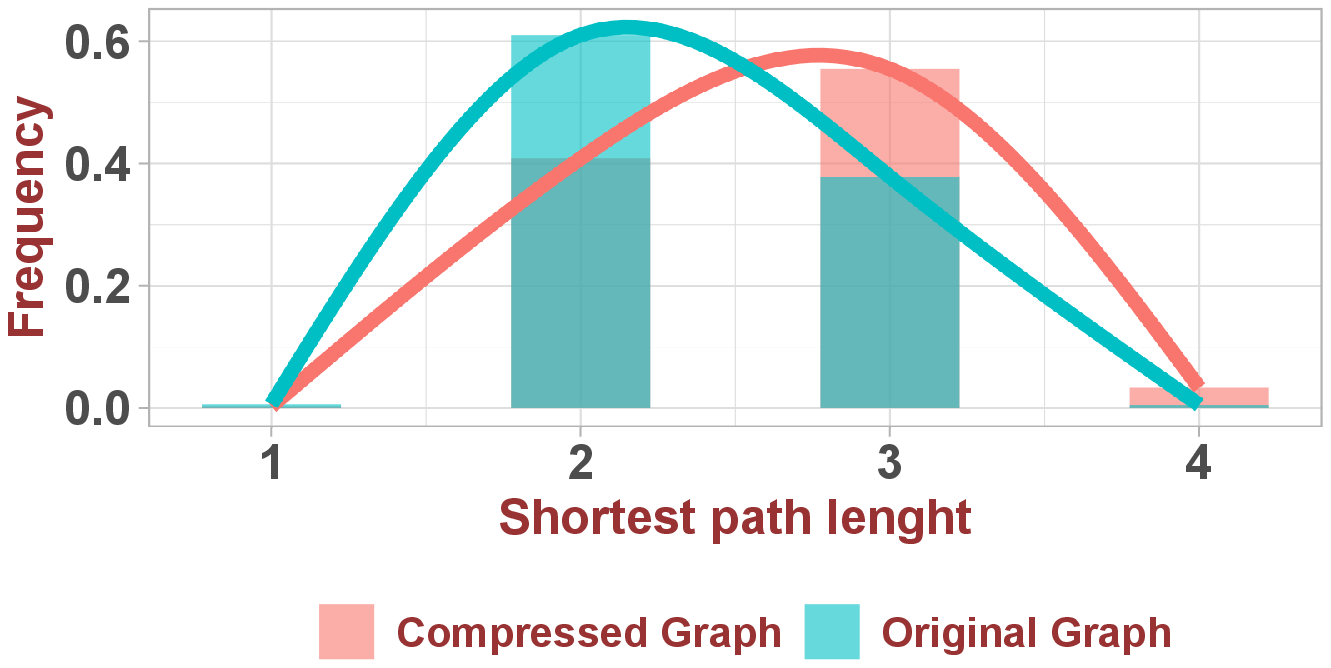}
    \caption{(b) Blog-Catalog dataset}
    \label{fig:3-b}
  \end{subfigure}
 \begin{subfigure}[b]{.49\textwidth}
    \includegraphics[width=1\textwidth, keepaspectratio]{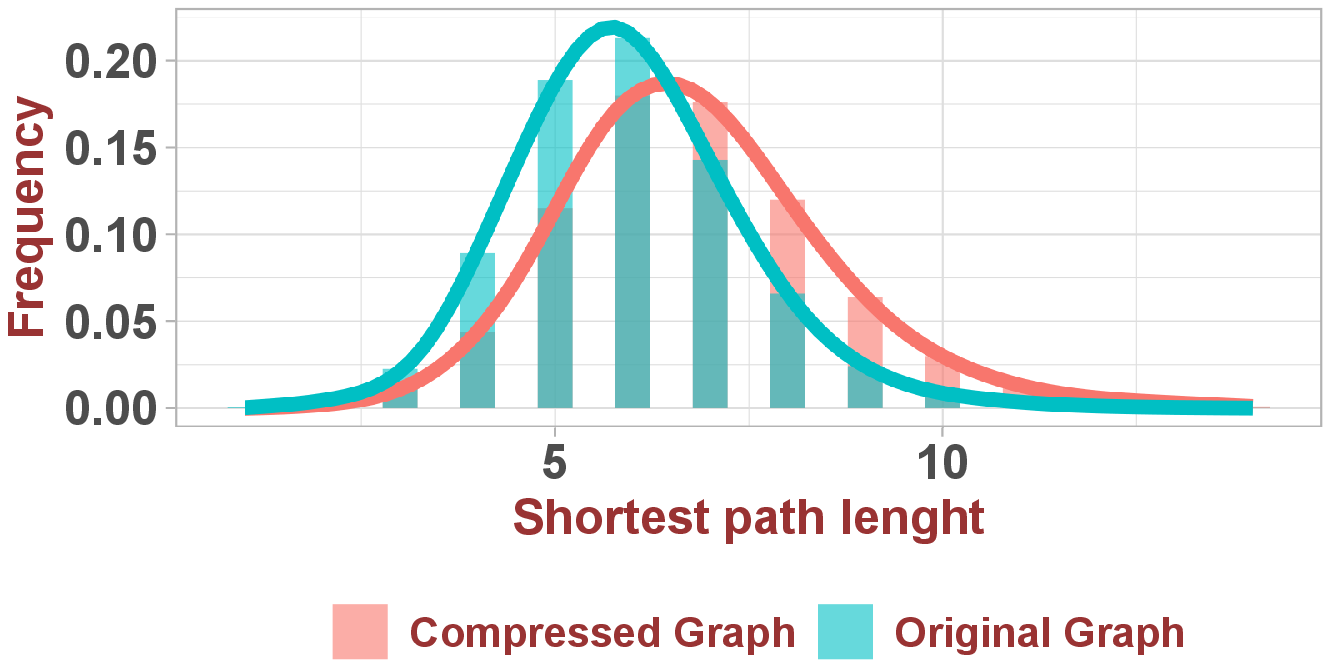}
    \caption{(c) Ca-HepTh dataset}
    \label{fig:3-c}
  \end{subfigure}
  \end{center}
  \caption{Distribution of shortest path lengths of the original and compressed graphs.  }
  \label{fig:Shortest_paths}
\end{figure}

Figure \ref{fig:Shortest_paths} shows the  distribution of the shortest path lengths in the original and compressed graphs for the three datasets. We note that the two curves have almost the same pace. This shows that our compression preserves the distribution of the lengths of the shortest paths in the 3 datasets. However, the curves of the compressed graphs are slightly stretched and shifted from the original curves. This is due to the stretching of the paths as a result of compression. This stretch is not really considerable because of the preservation of $50\%$ of the direct neighbors of each vertex in the graph. In addition, unlike $t$-spanners, $(p,t)$-compression compression controls the shift between the two curves by adjusting the value of parameter $p$.

\subsubsection{Whole graph embedding speed-up}
Many whole graph embedding methods are based on the local neighborhood information of the nodes. These  methods learn graph representations by exploring the node neighborhood and extracting some features such as walks, shortest paths, and local substructures. Since our compression preserves the local neighborhood within radius $t$, it is thus worth running these algorithms on compressed graphs to see if our compression speeds up these algorithms and to evaluate the performance loss. For this, we compressed three different datasets COLLAB, ENZYMES, and PROTEINS where $t=3$, $p(x<t)=0.0$ and $p(t)=1.0$ and we run three graph embedding algorithms on the compressed graphs: shortest path delta kernel \cite{borgwardt2005shortest}, graphlet kernel \cite{shervashidze2009efficient} and Graph2vec \cite{narayanan2017graph2vec}. We evaluated the performance of these algorithms on both the original and the compressed graphs in  graph classification tasks as follows: we train an SVM classifier with $90\%$ of the graphs  chosen randomly and then compute classification accuracy   on the test set composed of the remaining $10\%$ graphs. For Graph2vec, we used the best configuration of parameters given in the original paper \cite{narayanan2017graph2vec}.
\begin{table}[ht]
\centering
\caption{Graph kernel performance on the compressed graphs\\
cr: compression ratio.  pl: performance loss\\}
\begin{tabular}{ccccccc}
\hline
Dataset                   & Kernel                &cr  & Speed-up                    & Original accuracy                 & Accuracy & pl   \\ \hline
\multirow{3}{*}{COLLAB}   & SP delta  & \multirow{3}{*}{$79\%$}         & $4.52$                        & $65.78\% $                          & $64.78\%$  & $1.00\%$  \\
                          & 3-Graphlet          &                               & $9.39$                        & $64.62\%$                           & $53.80\%$  & $10.82\%$ \\
                          & 4-Graphlet          &                               & $\textgreater 10$ & \multicolumn{1}{l}{(out of time)} & $53.68\%$  & -       \\ \hline
\multirow{3}{*}{PROTEINS} & SP delta & \multirow{3}{*}{$40\%$}         & $1.123$                       & $71.91\%$                           & $72.18\%$  & $0.0\%$  \\
                          & 3-Graphlet      &                               & $1.49$                        & $71.60\%$                           & $71.18\%$  & $0.42\%$  \\
                          & 4-Graphlet          &                               & $2.02$                        &
                        $71.58\%$                           & $71.35\%$  & $0.23\%$   \\ \hline
\multirow{3}{*}{ENZYMES}  & SP  delta & \multirow{3}{*}{$39\%$}         & $1.06$                        & $29.31\%$                           & $24.71\%$  & $4.60\%$  \\
                          & 3-Graphlet           &                               & $1.6$                         &
                        $24.58\%$                          & $19.58\%$  & $5.00\%$ \\
                          & 4-Graphlet           &                               & $1.87$                        & $30.03\%$                           & $19.70\%$  & $10.33\%$ \\ \hline
\end{tabular}

\label{fig:graphKernels}
\end{table}

Table \ref{fig:graphKernels} shows the performance of the graph kernels on the compressed graphs. We notice that the three kernels run faster on compressed graphs in all experiments. This Kernel computation speed-up is more noticeable on  denser datasets, especially for the graphlet kernels. Indeed, we notice that the  4-Graphlet kernel exceeded the time limit (10 hours) on COLLAB's original graphs, while it takes less than one hour on the compressed dataset. Regarding performance loss, we notice a small loss in values for the shortest path kernels ranging from $0.0$ to $4.6\%$.
However, the loss is more noticeable for the graphlet kernel but it remains acceptable ($< 11.0\%$ in all experiments). This is due to the fact that our $(p,t)$-compression does not preserve all graphlets, for example only $6$ graphlets are preserved among the $11$ graphlets of size=4. Despite this,  
 kernel computation speed-up is very satisfactory.
\begin{table}[ht]
\centering
\caption{Graph2vec performance on compressed datasets}
\begin{tabular}{cccc}
\hline
\multicolumn{1}{l}{Dataset} & \multicolumn{1}{l}{Compression ratio} & \multicolumn{1}{l}{Speed-up} & \multicolumn{1}{l}{Performance   Loss} \\ \hline
COLLAB                      & $79\%$                                              & $2.54$                         & $0.00\%$                                 \\
PROTEINS                     & $40\%$                                              & $1.004$                        & $2.93\%$                                 \\
ENZYMES                     & $39\%$                                              & $1.27$                         & $5.99\%$                                 \\ \hline
\end{tabular}
\label{tab:graph2vecperf}
\end{table}

\begin{figure}[ht]
    \centering
     \includegraphics[width=0.50\textwidth, keepaspectratio]{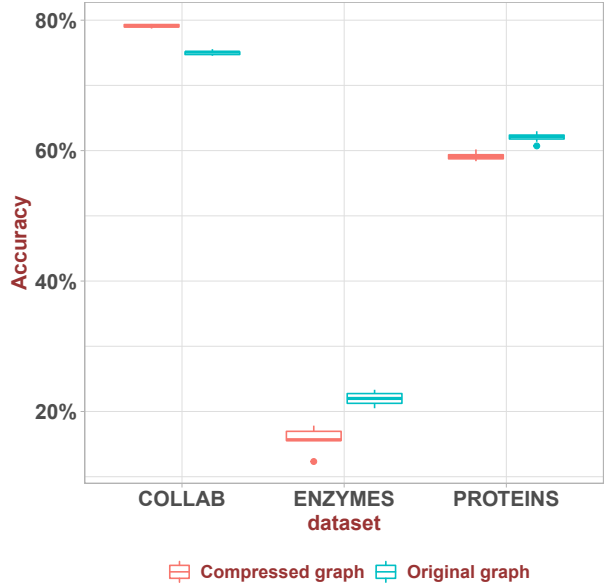}
     \caption{Graph2vec performance boxplot}
     \label{fig:graph2vec}
\end{figure}

Table \ref{tab:graph2vecperf} shows the performance of Graph2vec on the compressed datasets. Globally, the algorithm runs faster on the compressed graphs, and the speed-up factor is bigger on denser datasets. Performance loss is acceptable on the ENZYMES  dataset ($<6\%$) and very satisfactory on the first two datasets  ($< 3.0\%$). This is due to the fact that Graph2vec considers  graphs as sets of Weisfeiler-Lehman relabeled subgraphs \cite{shervashidze2011weisfeiler} that encompass higher order neighborhoods of graph nodes. These subgraphs are highly preserved 
 by our compression. Figure \ref{fig:graph2vec} depicts the distribution of the classification accuracy obtained using Graph2vec on the 3 datasets (compressed and original)  by running 10 experiments on each dataset and for each type of graph (compressed or original). We notice that the performances on compressed graphs and original graphs are nearly equivalent. The performance on original graphs is slightly better for the ENZYMES and PROTEINS datasets. Moreover, the loss of performance due to the compression is not great.

\subsubsection{Node embedding speed-up}
\begin{figure}[!h]
    \centering
     \includegraphics[width=0.80\textwidth, keepaspectratio]{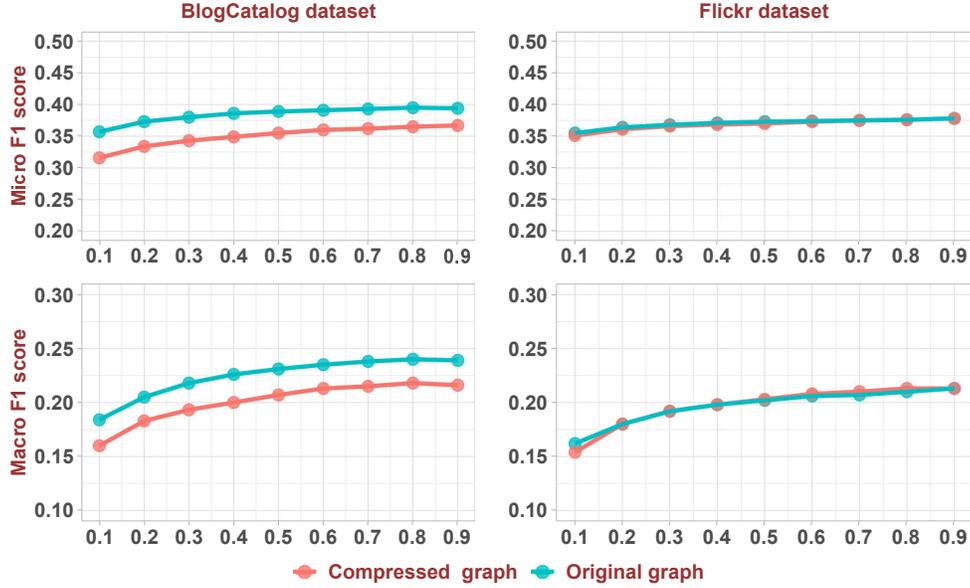}
      \caption{Performance of Node2vec on compressed graphs }
     \label{fig:n2v}
\end{figure}

\begin{figure}[!h]
    \centering
     \includegraphics[width=0.80\textwidth,keepaspectratio]{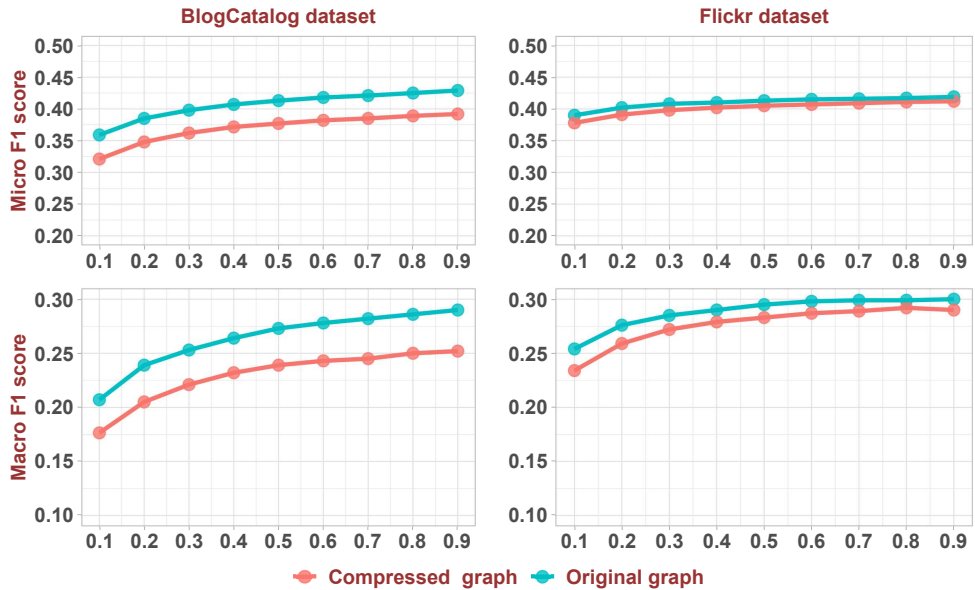}
      \caption{Performance of DeepWalk on compressed graphs}
     \label{fig:dw}
\end{figure}

In this series of experiments, we use two main algorithms, i.e.,  Node2vec  \cite{grover2016node2vec} and DeepWalk \cite{perozzi2014deepwalk},  to learn  node representations for both compressed and original graphs. We use BLOG-CATALOG and FLICKR datsets. The compressed graphs are obtained using $(p,t)$-compression where   $t=2$, $p(1)=0.5$ and $p(2)=1.0$. The compression ratios are $\geq 45\%$ for the two datasets. Node2vec and DeepWalk are run using the best parameter combinations given in the original papers. To evaluate their performance on both compressed and original graphs, we run  multiple multilabel classification tasks on the obtained representations. To this end, a sample $P_{tr} = \{0.1,...,0.9\}$ of the labeled nodes is used as training data. The rest of the nodes are used for  testing. This process is repeated $10$ times, and we use Macro-F1 and Micro-F1 as performance metrics. \\
\\
Table \ref{tab:node_emb} shows the performance of Node2vec and Deepwalk on the compressed graphs where $P_{tr}=0.5$. We notice that Deepwalk runs at the same speed on both the original and compressed graphs. This is because the time complexity of Deepwalk depends only on the number of nodes in the graph, which remains the same after our compression. However, Node2vec runs much faster  on the compressed graphs than on the original graphs. This is justified by the fact that Node2vec's time complexity depends on the square of the branching factor $b$ of the graph \cite{pimentel2018fast}, which is implicitly related to the number of edges in the graph. The Micro-F1 and Macro-F1 scores obtained on the compressed graphs are nearly  equivalent to the original scores. The performance loss rates are low ($\leq4\%$) for both methods, and insignificant on the FLICKR dataset ($\leq1.2\%$). \\
\\
For more fine-grained results, we also compared the performance of the two algorithms on compressed and original graphs while varying the size of the training sample $P_{tr}$  from $0.1$ to $0.9$. We summarize the results graphically for the Micro-F1 and Macro-F1 scores for both methods, i.e., Node2vec and Deepwalk, in Figures \ref{fig:n2v} and \ref{fig:dw}  respectively.  Here we make the same observations: the performances of the two algorithms on compressed graphs are almost similar to their performances on the original graphs for all training rates. With the  BLOG-CATALOG dataset, the performance drops by $4\%$ on compressed graphs in the worst case. However the performance curves on original and compressed graphs are almost identical in the FLICKR dataset.

\begin{table}[h]
\caption{Performance of node embedding algorithms  on compressed datasets}
\centering
\begin{tabular}{cccccc}
\hline
Dataset                      & Method    & Space gain               & Speed-up & Loss ( Micro F1) & Loss ( Macro F1) \\ \hline
\multirow{2}{*}{BlogCatalog} & DeepWalk & \multirow{2}{*}{$46.52\%$} & $0.99$     & $3.6\%$            & $3.4\%$            \\
                             & Node2vec  &                          & $2.87$     & $3.4\%$            & $2.5\%$           \\ \hline
\multirow{2}{*}{Flickr}      & DeepWalk & \multirow{2}{*}{$45.59\%$} & $0.99$     & $0.9\%$           & $1.2\%$            \\
                             & Node2vec  &                          & $5.32$     & $0.3\%$            & $0.0\%$            \\ \hline
\end{tabular}
\label{tab:node_emb}
\end{table}

\clearpage
\section{Conclusion and future work}
\label{Sec-Conclusion}
In this paper, we presented a graph summarization approach designed to control the amount of neighborhood information preserved in the computed summary. This approach relies on two parameters: a function $p$  that gives the proportion of each node's original neighbors to be preserved in its $i$-hops neighborhood in the compressed graph, and a threshold $t$ for which $p$ reaches its maximal value.

We presented algorithms to compute this compression with
the minimum cost, and showed their effectiveness in compressing input graphs through experimental evaluation
on multiple real life as well as synthetic graph datasets.

We also showed that the summaries computed by the proposed approach can be used without any decompression as input to  multiple graph applications, such as node embedding, graph classification, and shortest path approximations. The results show  interesting  trade-offs  between algorithm runtime speed-up and precision loss.

As for future work, we consider a more thorough analysis of $(p,t)$-compression impact on walks based graph learning algorithms such as Node2vec and DeepWalk. In fact, we observed some situations where learning accuracy increased when the graph was compressed. This was a quite unexpected observation. While we guess that walks are biased in the right direction by removing edges, characterizing such edges remains an open question. Another important open question is to find an efficient method to order graph edges. This would allow us to significantly improve the time complexity of the approach. In addition, we aim to design an incremental version of our compression to deal with dynamic graphs or graph streams.\\
 We note also that our approach can be used on both directed and undirected graph. However, our compression do not consider the labels of the edges. To compress edge-labelled graphs, a new model need to be defined so as to take into account these labels for example when defining the edge ordering.

\paragraph{ \textbf{Acknowledgement:}\\
This work is funded by ANR under grant ANR-20-CE23-0002.}

\bibliographystyle{unsrt}

\bibliography{references}

\begin{thebibliography}{10}

\bibitem{Liu2018}
Y.~Liu, T.~Safavi, A.~Dighe, and D.~Koutra.
\newblock Graph summarization methods and applications: A survey.
\newblock {\em ACM Comput. Surv.}, 51(3):62:1--62:34, June 2018.

\bibitem{Bloemena1976}
A.R. Bloemena.
\newblock {\em Sampling from a Graph}.
\newblock Mathematical Centre tracts. Mathematisch Centrum, 1976.

\bibitem{Zhang2017}
L.-C. Zhang and M.~Patone.
\newblock Graph sampling.
\newblock {\em METRON}, 75(3):277--299, Dec 2017.

\bibitem{Ahn2007}
Yong-Yeol Ahn, Seungyeop Han, Haewoon Kwak, Sue Moon, and Hawoong Jeong.
\newblock Analysis of topological characteristics of huge online social
  networking services.
\newblock In {\em Proceedings of the 16th International Conference on World
  Wide Web}, WWW '07, page 835–844, New York, NY, USA, 2007. Association for
  Computing Machinery.

\bibitem{Ravkic2018}
Irma Ravkic, Martin Žnidaršič, Jan Ramon, and Jesse Davis.
\newblock Graph sampling with applications to estimating the number of pattern
  embeddings and the parameters of a statistical relational model.
\newblock {\em Data Mining and Knowledge Discovery}, 32(4):913--948, July 2018.

\bibitem{Yousuf2020}
Muhammad~Irfan Yousuf and Suhyun Kim.
\newblock Guided sampling for large graphs.
\newblock {\em Data Mining and Knowledge Discovery}, 34(4):905--948, July 2020.

\bibitem{Nazi2015}
Azade Nazi, Zhuojie Zhou, Saravanan Thirumuruganathan, Nan Zhang, and Gautam
  Das.
\newblock Walk, not wait: Faster sampling over online social networks.
\newblock {\em Proc. VLDB Endow.}, 8(6):678–689, February 2015.

\bibitem{Li2019}
Y.~{Li}, Z.~{Wu}, S.~{Lin}, H.~{Xie}, M.~{Lv}, Y.~{Xu}, and J.~C.~S. {Lui}.
\newblock Walking with perception: Efficient random walk sampling via common
  neighbor awareness.
\newblock In {\em 2019 IEEE 35th International Conference on Data Engineering
  (ICDE)}, pages 962--973, 2019.

\bibitem{Zhao2019}
Junzhou Zhao, Pinghui Wang, John C.~S. Lui, Don Towsley, and Xiaohong Guan.
\newblock Sampling online social networks by random walk with indirect jumps.
\newblock {\em Data Mining and Knowledge Discovery}, 33(1):24--57, January
  2019.

\bibitem{Ribeiro2010}
Bruno Ribeiro and Don Towsley.
\newblock Estimating and sampling graphs with multidimensional random walks.
\newblock In {\em Proceedings of the 10th ACM SIGCOMM Conference on Internet
  Measurement}, IMC '10, page 390–403, New York, NY, USA, 2010. Association
  for Computing Machinery.

\bibitem{Riondato2016}
Matteo Riondato and Evgenios~M. Kornaropoulos.
\newblock Fast approximation of betweenness centrality through sampling.
\newblock {\em Data Mining and Knowledge Discovery}, 30(2):438--475, March
  2016.

\bibitem{Chew89}
L.~{Paul Chew}.
\newblock There are planar graphs almost as good as the complete graph.
\newblock {\em Journal of Computer and System Sciences}, 39(2):205 -- 219,
  1989.

\bibitem{Mathioudakis2011}
Michael Mathioudakis, Francesco Bonchi, Carlos Castillo, Aristides Gionis, and
  Antti Ukkonen.
\newblock Sparsification of influence networks.
\newblock In {\em Proceedings of the 17th ACM SIGKDD International Conference
  on Knowledge Discovery and Data Mining}, KDD '11, page 529–537, New York,
  NY, USA, 2011. Association for Computing Machinery.

\bibitem{Bonchi2013}
Francesco Bonchi, Gianmarco De~Francisci Morales, Aristides Gionis, and Antti
  Ukkonen.
\newblock Activity preserving graph simplification.
\newblock {\em Data Mining and Knowledge Discovery}, 27(3):321--343, November
  2013.

\bibitem{Nagano2011}
Kiyohito Nagano, Yoshinobu Kawahara, and Kazuyuki Aihara.
\newblock Size-constrained submodular minimization through minimum norm base.
\newblock In {\em Proceedings of the 28th International Conference on Machine
  Learning, ICML 2011}, Proceedings of the 28th International Conference on
  Machine Learning, ICML 2011, pages 977--984, 2011.
\newblock 28th International Conference on Machine Learning, ICML 2011 ;
  Conference date: 28-06-2011 Through 02-07-2011.

\bibitem{Brisaboa2009}
Nieves~R. Brisaboa, Susana Ladra, and Gonzalo Navarro.
\newblock k2-trees for compact web graph representation.
\newblock In Jussi Karlgren, Jorma Tarhio, and Heikki Hyyr{\"o}, editors, {\em
  String Processing and Information Retrieval}, pages 18--30, Berlin,
  Heidelberg, 2009. Springer Berlin Heidelberg.

\bibitem{Navlakha2008}
Saket Navlakha, Rajeev Rastogi, and Nisheeth Shrivastava.
\newblock Graph summarization with bounded error.
\newblock In {\em Proceedings of the 2008 ACM SIGMOD International Conference
  on Management of Data}, SIGMOD '08, page 419–432, New York, NY, USA, 2008.
  Association for Computing Machinery.

\bibitem{Rissanen1978}
J.~Rissanen.
\newblock Modeling by shortest data description.
\newblock {\em Automatica}, 14(5):465 -- 471, 1978.

\bibitem{Gallai1967}
T.~Gallai.
\newblock Transitiv orientierbare graphen.
\newblock {\em Acta Mathematica Hungarica}, 18:25--66, 1967.

\bibitem{Habib2010}
Michel Habib and Christophe Paul.
\newblock A survey of the algorithmic aspects of modular decomposition.
\newblock {\em Computer Science Review}, 4(1):41 -- 59, 2010.

\bibitem{Lagraa2016}
Sofiane Lagraa and Hamida Seba.
\newblock An efficient exact algorithm for triangle listing in large graphs.
\newblock {\em Data Mining and Knowledge Discovery}, 30(5):1350--1369,
  September 2016.

\bibitem{Suel2001}
T.~{Suel} and {Jun Yuan}.
\newblock Compressing the graph structure of the web.
\newblock In {\em Proceedings DCC 2001. Data Compression Conference}, pages
  213--222, 2001.

\bibitem{Boldi2004}
P.~Boldi and S.~Vigna.
\newblock The webgraph framework i: Compression techniques.
\newblock In {\em Proceedings of the 13th International Conference on World
  Wide Web}, WWW '04, page 595–602, New York, NY, USA, 2004. Association for
  Computing Machinery.

\bibitem{Koutra2015}
Danai Koutra, U~Kang, Jilles Vreeken, and Christos Faloutsos.
\newblock Summarizing and understanding large graphs.
\newblock {\em Stat. Anal. Data Min.}, 8(3):183–202, June 2015.

\bibitem{Neil2015}
Neil Shah, Danai Koutra, Tianmin Zou, Brian Gallagher, and Christos Faloutsos.
\newblock Timecrunch: Interpretable dynamic graph summarization.
\newblock KDD '15, page 1055–1064, New York, NY, USA, 2015. Association for
  Computing Machinery.

\bibitem{Maneth2018}
Sebastian Maneth and Fabian Peternek.
\newblock Grammar-based graph compression.
\newblock {\em Information Systems}, 76:19 -- 45, 2018.

\bibitem{Riondato2017}
Matteo Riondato, David García-Soriano, and Francesco Bonchi.
\newblock Graph summarization with quality guarantees.
\newblock {\em Data Mining and Knowledge Discovery}, 31(2):314--349, March
  2017.

\bibitem{Kumar2018}
K.~Ashwin Kumar and Petros Efstathopoulos.
\newblock Utility-driven graph summarization.
\newblock {\em Proc. VLDB Endow.}, 12(4):335–347, December 2018.

\bibitem{Shin2019}
Kijung Shin, Amol Ghoting, Myunghwan Kim, and Hema Raghavan.
\newblock Sweg: Lossless and lossy summarization of web-scale graphs.
\newblock In {\em The World Wide Web Conference}, WWW '19, page 1679–1690,
  New York, NY, USA, 2019. Association for Computing Machinery.

\bibitem{Fernandes2018}
Sofia Fernandes, Hadi Fanaee-T, and João Gama.
\newblock Dynamic graph summarization: a tensor decomposition approach.
\newblock {\em Data Mining and Knowledge Discovery}, 32(5):1397--1420,
  September 2018.

\bibitem{Amiri2018}
Sorour~E. Amiri, Liangzhe Chen, and B.~Aditya Prakash.
\newblock Efficiently summarizing attributed diffusion networks.
\newblock {\em Data Mining and Knowledge Discovery}, 32(5):1251--1274,
  September 2018.

\bibitem{Bloem2020}
Peter Bloem and Steven de~Rooij.
\newblock Large-scale network motif analysis using compression.
\newblock {\em Data Mining and Knowledge Discovery}, 34(5):1421--1453,
  September 2020.

\bibitem{Kapoor2020}
Sarang Kapoor, Dhish~Kumar Saxena, and Matthijs van Leeuwen.
\newblock Online summarization of dynamic graphs using subjective
  interestingness for sequential data.
\newblock {\em Data Mining and Knowledge Discovery}, September 2020.

\bibitem{Lee2020}
Kyuhan Lee, Hyeonsoo Jo, Jihoon Ko, Sungsu Lim, and Kijung Shin.
\newblock Ssumm: Sparse summarization of massive graphs.
\newblock In {\em Proceedings of the 26th ACM SIGKDD International Conference
  on Knowledge Discovery and Data Mining}, KDD '20, page 144–154, New York,
  NY, USA, 2020. Association for Computing Machinery.

\bibitem{zachary1977information}
Wayne~W Zachary.
\newblock An information flow model for conflict and fission in small groups.
\newblock {\em Journal of anthropological research}, 33(4):452--473, 1977.

\bibitem{Peleg1989}
David Peleg and Alejandro~A. Schäffer.
\newblock Graph spanners.
\newblock {\em Journal of Graph Theory}, 13(1):99--116, 1989.

\bibitem{CAI1994}
Leizhen Cai.
\newblock Np-completeness of minimum spanner problems.
\newblock {\em Discrete Applied Mathematics}, 48(2):187 -- 194, 1994.

\bibitem{Dinitz2016}
Michael Dinitz, Guy Kortsarz, and Ran Raz.
\newblock Label cover instances with large girth and the hardness of
  approximating basic k-spanner.
\newblock {\em ACM Trans. Algorithms}, 12(2), December 2016.

\bibitem{ELKIN2005}
Michael Elkin and David Peleg.
\newblock Approximating k-spanner problems for k>2.
\newblock {\em Theoretical Computer Science}, 337(1):249 -- 277, 2005.

\bibitem{KORTSARZ1994}
G.~Kortsarz and D.~Peleg.
\newblock Generating sparse 2-spanners.
\newblock {\em Journal of Algorithms}, 17(2):222 -- 236, 1994.

\bibitem{Gregory2008}
Steve Gregory.
\newblock A fast algorithm to find overlapping communities in networks.
\newblock In Walter Daelemans, Bart Goethals, and Katharina Morik, editors,
  {\em Machine Learning and Knowledge Discovery in Databases}, pages 408--423,
  Berlin, Heidelberg, 2008. Springer Berlin Heidelberg.

\bibitem{van1987simulated}
Peter~JM Van~Laarhoven and Emile~HL Aarts.
\newblock Simulated annealing.
\newblock In {\em Simulated annealing: Theory and applications}, pages 7--15.
  Springer, 1987.

\bibitem{borgwardt2005shortest}
Karsten~M Borgwardt and Hans-Peter Kriegel.
\newblock Shortest-path kernels on graphs.
\newblock In {\em Fifth IEEE international conference on data mining
  (ICDM'05)}, pages 8--pp. IEEE, 2005.

\bibitem{shervashidze2009efficient}
Nino Shervashidze, SVN Vishwanathan, Tobias Petri, Kurt Mehlhorn, and Karsten
  Borgwardt.
\newblock Efficient graphlet kernels for large graph comparison.
\newblock In {\em Artificial Intelligence and Statistics}, pages 488--495,
  2009.

\bibitem{narayanan2017graph2vec}
Annamalai Narayanan, Mahinthan Chandramohan, Rajasekar Venkatesan, Lihui Chen,
  Yang Liu, and Shantanu Jaiswal.
\newblock graph2vec: Learning distributed representations of graphs.
\newblock {\em arXiv preprint arXiv:1707.05005}, 2017.

\bibitem{shervashidze2011weisfeiler}
Nino Shervashidze, Pascal Schweitzer, Erik~Jan Van~Leeuwen, Kurt Mehlhorn, and
  Karsten~M Borgwardt.
\newblock Weisfeiler-lehman graph kernels.
\newblock {\em Journal of Machine Learning Research}, 12(9), 2011.

\bibitem{grover2016node2vec}
Aditya Grover and Jure Leskovec.
\newblock node2vec: Scalable feature learning for networks.
\newblock In {\em Proceedings of the 22nd ACM SIGKDD international conference
  on Knowledge discovery and data mining}, pages 855--864, 2016.

\bibitem{perozzi2014deepwalk}
Bryan Perozzi, Rami Al-Rfou, and Steven Skiena.
\newblock Deepwalk: Online learning of social representations.
\newblock In {\em Proceedings of the 20th ACM SIGKDD international conference
  on Knowledge discovery and data mining}, pages 701--710, 2014.

\bibitem{pimentel2018fast}
Tiago Pimentel, Adriano Veloso, and Nivio Ziviani.
\newblock Fast node embeddings: Learning ego-centric representations.
\newblock 2018.

\end{thebibliography}

\end{document}